# Extraction and Simulation of the Impact of Flux Trapping in Moats on AC-Biased Shift Registers

Scott E. Meninger and Sergey K. Tolpygo, *Senior Member, IEEE*

*Abstract*—Moats in superconducting ground planes are used to trap magnetic flux away from sensitive parts of superconductor integrated circuits. We simulate the effect of magnetic flux trapped in moats on the operating margins of ac-biased SFQ shift registers (ShReg) with two ground planes for various congruent moat geometries, moat sizes, locations in the ShReg cells, number and polarity of the trapped fluxons. Using inductance extractor InductEx, we extract mutual couplings between the moats and the ShReg inductors and include them in the refined netlist. Then we use JoSim to simulate the circuit operation and find the threshold ac clock amplitude above which the register starts to operate correctly. The relative change in this threshold is used to characterize the influence of flux trapping on the circuit operation.

Monte-Carlo simulations are used to investigate the effect of the circuit parameter variations on the statistics of the threshold, using the standard deviation of the Josephson junction critical current and of the circuit inductance extracted from the fabrication process control monitors. The obtained distributions are compared with the independently measured distributions of the threshold amplitude for the individual cells of an ac-biased shift registers with 108,500 Josephson junctions. The results show that flux trapping in properly designed moats have a very small effect on the operation margins of the large-scale shift registers, whereas the main contribution to the cell-to-cell variation of the threshold comes from the variation of the junctions' critical currents.

## I. INTRODUCTION

MOATS are employed in superconductor digital circuits (SDC) to preferentially trap flux during circuit cooldown. The size of the moats and their location can have an impact on circuit behavior when flux trapped in the moats couples to adjacent circuitry [1], [2]. In this work we look at the change in the operation margins due to trapped flux of a shift register circuit that was previously proposed as a convenient process monitoring tool [3]. Specifically, we look at the change in the ac clock positive amplitude threshold above which the shift register operates correctly, the so-called lower positive (PL) margin in notations of [3], and at the minimum negative amplitude of the ac clock above which the circuit operates correctly, the negative lower (NL) margin [3].

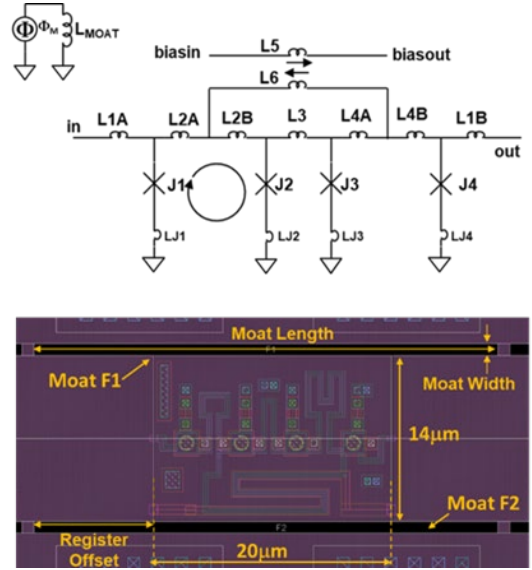

**Fig. 1**. AC-biased shift register cell schematics incorporating inductive coupling to a ground plane moat modeled as a flux source (top). Bottom panel – layout of a single register cell, showing two moats, F1 and F2, congruent in the M4 and M7 ground planes [4].

The shift register cell design and operation were described in [3], [4]. Briefly, upon reaching the PL threshold of the clock amplitude, the data "1" (flux quantum) moves inside the cell from its first part to its second part, where it is stored until the negative clock polarity is applied. When the negative clock amplitude reaches the NL threshold value, the data fluxon is shifted to the next cell (into its first part) in the register, and so on.

We utilize the InductEx [5] extraction toolset and simulate the circuit with JoSim [6]. A similar approach was used in [2], [7]. In order to validate the simulations, we compare our results with measured data. The incorporation of Monte Carlo modeling to account for process variation yields a reasonably good match to the main distribution of the measured thresholds of the individual cells in the registers containing more than 27,000 cells and 108,500 Josephson junctions. We then expand our model to include the possibility that some outliers lying far outside the main data distribution (sometimes occurring randomly upon thermal cycling) may be caused by Abrikosov vortices trapped in superconducting films elsewhere in the

This material is based upon work supported under the Air Force Contract No. FA8702-15-D-0001.

(Corresponding author: Sergey K. Tolpygo, e-mail: sergey.tolpygo@ll.mit.edu).

Both authors are with the Lincoln Laboratory, Massachusetts Institute of Technology, Lexington, MA 02421 USA (e-mails: scott.meninger@ll.mit.edu, sergey.tolpygo@ll.mit.edu ).

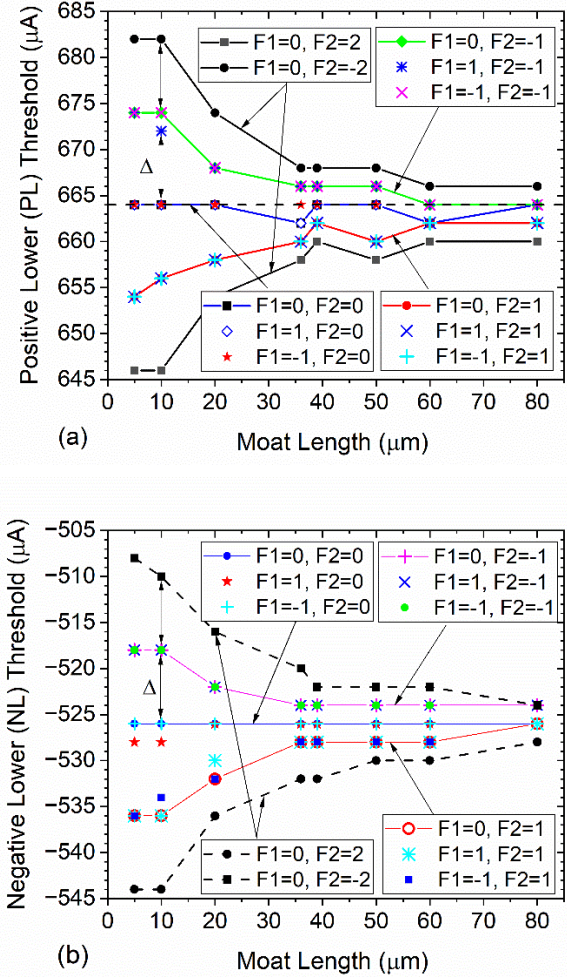

**Fig. 2**. AC clock amplitude thresholds of the register cell vs moat length: a) positive lower (PL) threshold; b) negative lower (NL) threshold. The moat width $w$ is 1 µm.

TABLE I
SIMULATED EFFECT OF FLUX TRAPPING IN MOATS

| Moat length × width (µm × µm) | Pitch, $p_x$ (µm) | Cell distance to moat edge (µm) | PL (mean), no fluxons (µA) | NL (mean), no fluxons (µA) | PL [min, max] ±1Φ₀ in moats (µA) | NL [min, max] ±1Φ₀ in moats (µA) |
|---|---|---|---|---|---|---|
| 3 × 3 | 10 | 8.5 | 664 | -526 | [658, 668] | [-532, -520] |
| 4 × 4 | 10 | 7.5 | 662 | -524 | [658, 666] | [-528, -520] |
| 5 × 5 | 10 | 6.5 | 664 | -526 | [660, 666] | [-530, -524] |
| 4 × 4 | 20 | 0 | 666 | -526 | [664, 666] | [-526, -526] |
| 5 × 5 | 20 | 0 | 664 | -526 | [664, 664] | [-526, -526] |
| 36 × 0.3 | 40 | 0 | 664 | -526 | [662, 664] | [-526, -526] |

circuit outside the moats, e.g., near defects in the superconducting Nb films.

## II. FLUX TRAPPING IN MOATS

To investigate the impact of trapped flux on the performance of SDC we utilize the ac-biased shift register cell design presented in [3], [4]. This single flux quantum (SFQ) shuttle is depicted in Fig. 1 and uses an ac bias to move a unit flux through its structure as follows. A fluxon that initially enters via the input and gets stored in the loop comprising J1 and J2 will move to the loop formed by Josephson junctions (JJs) J3 and J4 (into the second part of the cell) when the positive bias current coupled from the ac bias (clock) line exceeds some positive value, PL. Under this condition, the J3-J4 loop is negatively biased, so the fluxon does not move further. When the bias current reverses polarity, the J3-J4 loop becomes positively biased, and the fluxon moves to the next cell connected to the output when the negative amplitude research the NL threshold value, and a shift occurs. By varying the magnitude of the current coupled to the register, we can determine the positive and negative thresholds of the register. In Fig. 1, $\Phi_M$ represents flux that is captured by a given moat. $L_{MOAT}$ is the moat inductance that couples to the various circuit inductances through mutual inductances, which are not shown in the schematic for simplicity. Each of the moats has its own moat inductance, flux, and mutual inductances.

Also shown in Fig. 1 is the layout of the register cell, which was fabricated in the Lincoln Laboratory SFQ5ee process with a nominal critical current density $J_c$=100 µA/µm² [8]. The moats, labeled F1 and F2 in the layout are created by connecting the slits in sky and ground planes with bridging metal. Moat length, $l$ along the in Fig. 1, width $w$, and pitch $p_x$ – the moat placement period along the $x$-direction – are varied for the 20 µm long and $W$ =14 µm wide register cell. The $y$-direction moat placement pitch is $p_y = W + w$. The moat F2 is spaced 0.9 µm from the centerline of the ac bias transformer L5-L6.

### A. Extraction and Simulation Results

Fig. 2 shows the results of an extraction and simulation for the register cell with the moat width $w$ =1 µm. The moat length is varied from 5 µm to 96 µm. The positive lower (PL) and negative lower (NL) thresholds are reported. These represent the clock threshold amplitudes above which, in absolute value, fluxons will propogate through each stage of the register on, respecitvely, the positive and negative half-periods of the ac clock, consistent with the methodology described in [3]. The number and polarity of fluxons in the moats are represented as 0,±1,±2, etc.

In the presentation hereafter, we will concentrate on the effects of the singly occupied moats − moats containing a single flux quantum – mainly to save space because the number of possible combinations grows rapidly with increasing the number of fluxons. To estimate the effect of multiple fluxons in the moats, the single-fluxon results presented hereafter can be simply scaled due to the superposition principle: the mutual inductances between the moat and the inductors do not depend on the number of fluxons in the moat, while the screening current circulating the moat and the change in the operating thresholds grow proportionally to the number of the fluxons. This proportionality can be seen in the simulated change Δ of the PL and NL thresholds with increasing the number of fluxons in the moat F2 from zero to ±2 while keeping the number of fluxons in the moat F1 fixed, e.g., at zero; see Fig. 2.

In the considered circuit and moat designs, and at the typical experimental conditions [9], [10], the probability of trapping multiple flux quanta in the moat is low due to the following. A vortex attracts to a moat; however, it repels from a fluxon of the same polarity already trapped in the moat. Filling the moat by vortices may proceed only if the attraction force is larger than the repulsion force. More precisely, moat flux filling may proceed if the free energy of the ground plane film with the vortex is larger than the free energy of the same film without the vortex but with an extra fluxon in the moat. In the typical range of residual magnetic fields in the mu-metal shields used in superconductor digital circuit testing, $B_r \sim$ 0.1 μT to 1 μT [9], [10], the number density of vortices in the film $n_v \approx B_r/\Phi_0$ is between about $5 \cdot 10^3$ and $5 \cdot 10^4$ cm$^{-2}$. The average distance between them $a \sim (\Phi_0/B_r)^{1/2}$ is large, about 45 μm to 144 μm at the ends of the field range. Therefore, if the number density of the moats $n_{moat}$ in the circuit design is much larger than the expected vortex number density, the probability of finding two or more vortices near the same moat is very low. For the regularly spaced moats, as in the design considered here, the latter condition is equivalent to having the moat placement pitches in x- and y-directions $p_x, p_y \leq a$. In the shift register in Fig. 1, $p_y =$ 15 μm and $p_x$ was varied up to 100 μm [10]; see Fig. 2. Therefore, if the condition $n_{moat} > n_v$ is satisfied, we can limit the analysis to the singly occupied moats.

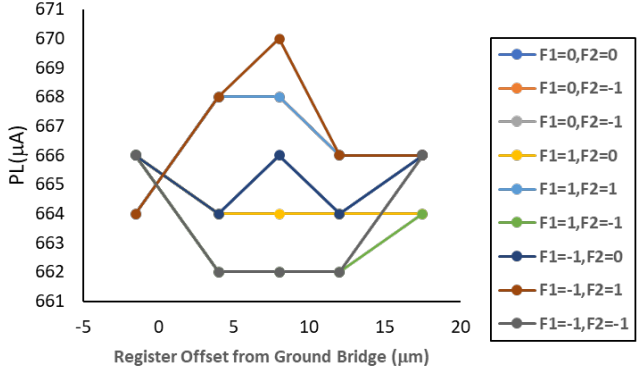

Fig. 3. Effect of cell offset with respect to the moat edge on the positive lower (PL) threshold; the moat $l \times w =$ 5 μm x 5 μm, the moat placement pitch along the cell $p_x =$ 20 μm.

TABLE II
MEAN VALUES AND STANDARD DEVIATIONS OF SIMULATED WITH MONTE-CARLO AND MEASURED THRESHOLDS

| Parameter | Measured (μA) | Extraction and Monte-Carlo Simulations (μA) | (Simulated −Measured)/Measured (%) |
|---|---|---|---|
| PL threshold | 639 | 636 | <1 |
| 1-σ PL | 15.3 | 11.7 | −23.5 |
| NL threshold | 526 | 489 | 7 |
| 1-σ NL | 15.7 | 13.2 | −18.9 |

We make the following obervations for the various combinations of fluxons. The change in the threshold from the zero-fluxon nominal case (PL=664 μA, NL=−524 μA) is larger for shorter moats than for longer moats. For longer moats, longer than about 30 μm, the effect on the thresholds diminishes. These observations are consistent with the moat inductance increasing as moat length increases, reducing the screening current circulating around the moat. Also, the mutual inductance between the moat and the key parts of the circuit no longer increases with the moat inductance when the moat length becomes larger that about the cell size. Finally, we note that the absolute value and the relative change in the thresholds are small in the comparison to the thresholds; even for the smallest moats the miximum change is ~3% for PL and 3.4% for NU.

Table I presents a summary of the simulated effect of fluxons trapped in the moat of various configuraions on the PU and NU thresholds of the shift resiter cell. For all the considered cases, the maximum and minimum value of PL and NU are reported for 0, 1, and −1 fluxons present in moats F1 and F2. The maximum deviation is very small compared to the absolute value of either threshold.

*B. Cell Offset Sweep Result*

Table I shows multiple cases for 4 μm × 4 μm and 5 μm × 5 μm moats, differing by the offset of the moat with respect to the regsiter cell. Offset is annotated in the layout presented in Fig. 1 as the distance from the cell edge to the leftmost edge of the moat (distance to the bridge in the ground plane). Changes in the moat offset changes its coupling to the circuit and, therefore, changes the operation thresholds. These simulations were carried out by changing the moat offset in the layout. This is clearly observable as the shape of the curve in Fig 3. Maximum coupling occurs when the moat is offset to align with the center of the cell, and the minima occurs when the moat is offset to align to the edges of the cell. This plot is presented for 5 μm × 5 μm moats on a 20 μm pitch for the 20-μm-long register cell to most clearly demonstrate the behavior, since the pitches will align as the offset is varied. We performed simulations for different pitches and observed the same order of magnitude peak deviations. Figure 3 demonstrates that the net impact of the cell-to-moat offset is small across all different offsets, representing no more than ~1.2% of the nominal threshold value.

III. ADDING IN VARIABILITY OF PROCESS PARAMETERS

In order to validate the extraction and simulation flow, we compared the simulations to experimental data in [10]. Figure 4 presents experimental data measured on a long shift register with 4513 cells using long rectangular moats with $l \times w =$ 36 μm × 1 μm. Each moat has two cells aligned on one side of the moat and two mirror-inverted cells (belonging to the adjacent row) symmetrically aligned on the other side of the moat. The main body of the distributions for both the positive and negative thresholds show much greater variation than is explained by our simulations of the effect of the trapped fluxons in the moats.

To account for process parameter variation, Monte-Carlo variation was added to the simulations by introducing normal distributions for critical currents of Josephson junctions $I_c$ (1σ= 1.3%) and inductance (1σ=0.6%), taken from the process control monitor data measured independently and described in [11], [12], [13], [14]. In addition, the process mean critical current density, $J_c$ was set to be slightly lower ($J_{c-\mu} =$

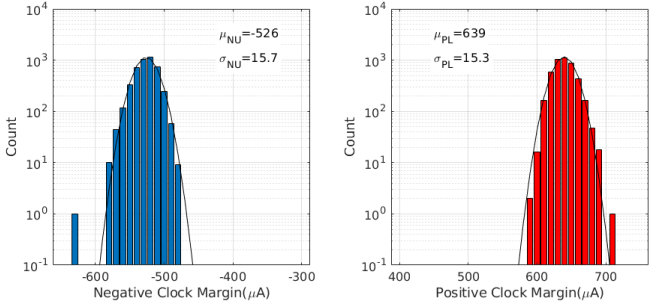
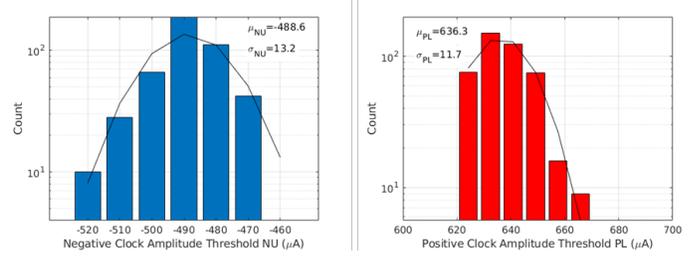

**Fig 4.** Measured distributions of the positive lower (PL) and negative lower (NL) thresholds (margins) for shift "1" operation of the individual cells in a 4513-bit shift register (20,052 JJs) with long moats, $l \times w = 36\ \mu m \times 1\ \mu m$, and 4 μm ground bridges [10]. Two 20-μm-long cells are aligned on each side of each moat in the register; the moat number density $n_{moat} \approx 1.67 \cdot 10^5\ cm^{-2}$. The mean value and the standard deviation of the distributions are given in the legend. The special U-turn cells having a slightly different design and threshold values were filtered out from the distributions.

**Fig 5.** Simulated distributions of the positing lower and negative upper thresholds for shift "1" operation of the shift register cells with account for Monte Carlo variation of the junctions' critical current (1σ=1.3%) and cell inductors (1σ=0.6%). Solid lines are fits to Gaussian distributions. The mean values and the standard deviations nearly match those extracted from the measurements in Fig. 4.

95 μA/μm²) to match the measured $J_c$ at the location of the shift register chip in Fig. 4. This made the simulations equivalent to a process corner run.

Figure 5 shows the results of these simulations and the fits to a Gaussian distribution. Table II gives comparison of the simulated and the measured distributions of the cell thresholds.

There is a very good agreement between the mean values of the both thresholds. However, the simulated standard deviations (widths of the threshold distributions) are systematically smaller than in the measurements. This is easily explainable by the fact that the measured distributions include a contribution from the measurement error in the threshold determination, which can be characterized by a standard deviation $\sigma_{meas}$ and estimated as $\sigma_{meas} \approx 10$ μA. This error is caused by the finite digitization step corresponding to the least significant bit in the ac clock source, and thermal and environmental noise. Then, the total variance of measured distribution is $(\sigma_{fab}^2 + \sigma_{flux}^2 + \sigma_{meas}^2)$, where $\sigma_{fab}^2$ is the variance caused by the fabrication process and $\sigma_{flux}^2$ is the variance cause by random flux trapping in the moat. The latter is at most a few μA² according to the data in Sec. IIA and thus can be neglected. Then, from the data in Table II for the PL threshold, the $\sigma_{fab} \approx (15.3^2 - 10^2)^{\frac{1}{2}} = 11.6$ μA. Similar estimate for the NL threshold gives $\sigma_{fab} \approx 12.1$ μA. These values agree nicely with the Monte-Carlo simulated distribution widths in Table II.

Also, the shift registers consist of two distinct groups of cells – those in the rows and the so-called U-turn cells connecting the ends of the rows; see [3]. The layout of the U-turn cells is different due to the difference between the cell length in the rows (in the $x$-direction) and the distance between the rows (in the $y$-direction). As a result, U-turn cells have a different (lower) mean values of the PL than the cells in the rows [10]. In the distribution shown in Fig. 4, we filtered out all U-turn cells. In addition, there are a few cells with special locations adjacent to the input DC-to-SFQ converter and to the output SFQ-to-DC drivers. DC bias of those Input/Output cells of the shift register affect the operating thresholds of the adjacent bit cells, making them look like outliers in Fig. 4, while they are not because they are repeatable across thermal cycling and have consistent locations in the register chain. These special cells make the measured distribution slightly wider than the actual distribution of the cells in the rows which were used in the Monte-Carlo simulations.

Despite this very good agreement, in some rare cases, we observed some outlier cells varying their threshold significantly across thermal cycling and lying outside the main distribution statistics [10]. At the same time, positions of these cells in the register do not change with thermal cycling. It is reasonable to assume that these rarely occurring outlier cells have some structural defects, e.g., in the ground plane film(s), causing flux trapping inside the cell rather than in the moat. We next seek to capture these outliers in our simulation model.

### III. ADDING ABRIKOSOV VORTICES

Small openings of 0.1 μm x 0.1 μm were inserted congruently in the ground planes layout to model the presence of Abrikosov vortices in the ground planes. These dimensions roughly correspond to the value of magnetic field penetration depth, λ=90 nm, in Nb films used in the SFQ5ee process at MIT LL [12], [13], [14]. The moat extraction and simulation flow was run to determine what the impact of a single flux quantum $\Phi_0$ placed in the opening would be when placed at various points throughout the main circuit area, as depicted in Fig 6. In [1] an energy-based analysis is presented and described to determine where vortices may be more likely to occur. For this work we decided to simply place a few model vortices around the layout to show that we see a threshold shift on the same order as what we observe experimentally in the rare outlier cells. This does not suggest that we know the exact location of the vortex, but that the value of the threshold shift could correspond to a vortex.

As the first row of Table III shows, for a number of vortex locations placed inside the circuit area, the shift in PL is not very significant, moving only 20 μA peak to peak across all fluxon combinations. This represents 3% of the nominal threshold value. On the other hand, once a vortex is placed close to a Josephson junction (location F23) then, as expected, the threshold shift is much larger. Rows 3-4, 6-7, and 9-10, marked in the Table III, demonstrate that vortex F23 causes a very significant shift in PL of about 400 µA peak-to-peak depending on the fluxon sign, or 60% of the nominal threshold.

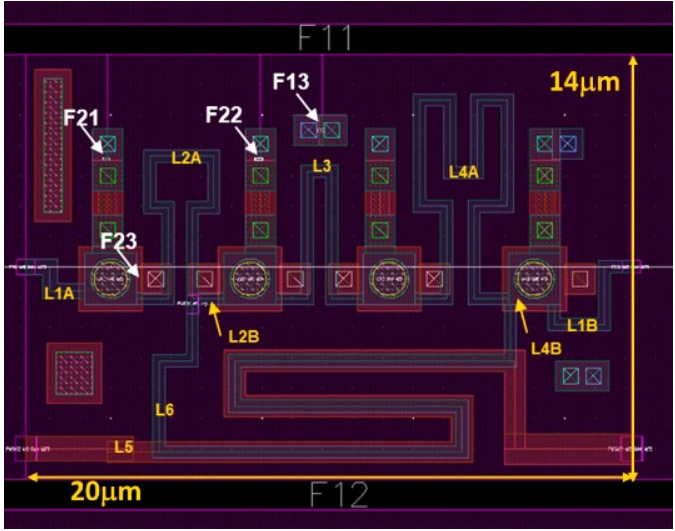

Fig 6. Modified layout of the register unit cell including small congruent openings, shown as F21, F23, F22, and F13, in the M4 and M7 ground planes modeling various potential locations of Abrikosov vortices in the cell. The cell length and the ground plane track width $W$ between the moats are also shown.

TABLE III
SIMULATED SHIFTS OF THE PL THRESHOLD AT VARIOUS POSITIONS OF THE MODEL VORTEX IN THE CELL

| F13 | F21 | F22 | F23 | PL (µA) |
|---|---|---|---|---|
| 0,1,-1 | 0,1,-1 | 0,1,-1 | 0 | 654 → 674 |
| 0 | 0 | 0 | 0 | 670 |
| 0 | 0 | 0 | 1 | 470 |
| 0 | 0 | 0 | -1 | 870 |
| 1 | 0 | 0 | 0 | 680 |
| 1 | 0 | 0 | 1 | 480 |
| 1 | 0 | 0 | -1 | 880 |
| -1 | 0 | 0 | 0 | 660 |
| -1 | 0 | 0 | 1 | 470 |
| -1 | 0 | 0 | -1 | 860 |

## IV. CONCLUSIONS AND FUTURE WORK

We have demonstrated through testing that moats are capable of trapping flux and protecting circuits for different moat constructions. Further, the coupling from the trapped flux to the adjacent circuits have a negligible impact on the circuit operation as the majority of the variability in the circuit margin comes from process parameter spreads. Outliers that change register location over thermal cycling are most likely caused by vortices trapped in the film outside the moat elsewhere in the circuit. Accounting for the possibility of these vortices requires additional steps in pre-fabrication analysis including the impact of trapped flux on $I_c$ of Josephson junctions and the use of energy-based analysis to predict the location of the vortices.

Experimental data in [10] show that slit-like moats of the minimum processed-allowed width, e.g., $w = 0.3$ µm, and length $l \gg w$, having negligible effect on the overall circuit density, protect circuits from flux trapping outside of the moats in 100% of cooldowns when $n_{moat} > n_v$. The latter condition overestimates the required number of the moats but allows one to avoid making complex free energy calculations of whether the moats can be filled by multiple fluxons.

Another important consideration in designing moats is the moat placement pitch in the $y$-direction, i.e., the effective width $W$ of the ground plane track between the moats; see Fig. 6. It should be based on the value of residual magnetic field in the circuit test setup $B_r$. To allow for the full expulsion of vortices from the ground planes, the maximum width of the film track between the moats should be less than $W < W_K = (1.65\Phi_0/B_r)^{1/2}$, that follows from the flux expulsion from a single superconducting strip [15]. This estimate works well for long slit-like moats comparable in length with the cell length [10]. Decreasing the moat length at a fixed moat pitch reduces the numerical coefficient in the formula for $W_K$. Numerical simulations are required for other moat shapes and arbitrary placement of moats. We are not aware of such simulations.

The presented study was done for the congruent moats in the top (layer M7) and bottom (layer M4) niobium ground planes sandwiching the circuit; the ground planes are 1.4 µm apart and connected by superconducting vias [8]. Experiments [10] show enhancement of flux trapping in the films if the moats are noncongruent and a strong dependence of the flux trapping on the number of the ground planes and their vertical separation. These three-dimensional (3D) effects in the interaction of vortices on different superconducting layers cannot be simulated using the existing tools [5]. Modeling and analyzing the 3D effects in a stack of multiple superconducting layers and the inter-layer flux trapping caused by a parallel or inclined magnetic field require further work and development of the software and simulation tools.

## ACKNOWLEDGMENT

We are grateful to Evan Golden and Neel Parmar for sharing with us the experimental data on the distributions of the operation thresholds in various ac-biased shift registers they tested and for many useful discussions. We are thankful to Vasili Semenov who designed and optimized the ac-biased shift register utilized in this work. Discussions of flux trapping with him are also gratefully acknowledged. We thank Vladimir Bolkhovsky and Ravi Rastogi for overseeing the wafer fabrication runs of the SFQ5ee process used to make the shift registers.